 \documentstyle[prd,aps,preprint]{revtex}
 \draft
 \tighten

\begin{document}
 \preprint{hep-ph/9503438}

 \title{Screening of Long-Range Leptonic Forces by Cosmic Background
 Neutrinos}
 \author{Alexandre D.~Dolgov$^*$}
 \address{Dept.\ de Fisica Teorica, Universidad de Valencia,
 46100 Burjasot, Spain}
 \author{Georg G.~Raffelt}
 \address{Max-Planck-Institut f\"ur Physik,
 F\"ohringer Ring 6, 80805 M\"unchen, Germany}
 \date{\today}
 \maketitle
 \begin{abstract}
 The absence of dispersion effects of the SN~1987A neutrino pulse has
 been used to constrain novel long-range forces between neutrinos and
 galactic baryonic or non-baryonic matter. If these forces are
 mediated by vector bosons, screening effects by the cosmic neutrino
 background invalidate the SN~1987A limits and other related
 arguments.
 \end{abstract}
 \pacs{PACS number(s): 12.60.--i, 13.15.+g, 95.35.+d, 97.60.Bw}


 The $\overline\nu_e$ signal observed from the supernova (SN) 1987A
 lasted for a few seconds \cite{SN} so that one can derive limits on
 a variety of effects that would have caused a significant
 dispersion.  The most thoroughly studied case is that of an assumed
 neutrino mass which would introduce a time delay relative to
 massless neutrinos of $\delta t/t=m^2_\nu/2E_\nu^2$ \cite{Zatsepin}.
 With a $\delta t$ of a few seconds, the time of flight to the SN of
 $t=50\,{\rm kpc}/c\approx5{\times}10^{12}\,\rm s$, and energies
 spread over several $10\,\rm MeV$ one arrives at a mass limit
 $m_\nu\alt20\,\rm eV$. Another interesting case is the limit on a
 putative neutrino charge because the motion on curved paths in the
 galactic magnetic field would also lead to a dispersion
 $\delta t/t\propto E_\nu^{-2}$ \cite{Cocconi}. Yet another case is
 an assumed energy-dependent velocity of massless particles due to a
 fundamental length scale~\cite{Fujiwara}.

 We presently take a second look at attempts to limit novel
 long-range forces which would bend the neutrino trajectories and
 thus also lead to a dispersion $\delta t/t\propto E_\nu^{-2}$
 \cite{FM}.  The force was assumed to be mediated by massless vector
 bosons coupled to a novel neutrino charge $q_\nu$ which may
 represent, for example, a leptonic interaction. The source of the
 force are the electrons of the galaxy, or perhaps the protons or
 dark matter particles with a charge $q_e$, $q_p$, or $q_x$.
 Depending on the nature of the source particles and their
 distribution in the Milky Way, a limit
 \begin{equation}\label{E1}
 |q_{e,p,x} q_\nu|\alt3{\times}10^{-40} \times
 \cases{1&for $e$, $p$\cr m_x/m_p&for dark matter\cr}
 \end{equation}
 was derived \cite{FM}. For dark matter particles the mass $m_x$
 appears because the mass density, not the number density of dark
 matter particles, is fixed by the galactic rotation curves.

 By assumption, in this scenario neutrinos carry a novel charge
 $q_\nu$ which must be opposite between $\nu_e$ and $\overline\nu_e$
 as in normal electromagnetism. Because the cosmic neutrino
 background is likely CP~symmetric it constitutes a neutral plasma
 with regard to the new charge and interaction. Therefore, one
 expects that a source for the new force such as a star or a galaxy
 will be screened according to the standard Debye-H\"uckel theory and
 the bound~(\ref{E1}) will be invalid. Similarly, screening effects
 will invalidate the time delay argument between $\nu_e$'s and
 $\overline\nu_e$'s discussed in Ref.~\cite{PSW}.

 A very strong bound on a leptonic charge coupling,
 $\alpha_L\equiv q_L^2/4\pi\alt 10^{-49}$,
 follows from high-precision tests of the equivalence principle
 \cite{lbo}. At first sight this limit is also invalidated by the
 possible screening due to background neutrinos~\cite{screen} so that
 the strongest bound would come from $\nu_e e$-scattering,
 $\alpha_L \alt 10^{-12}$. However, a more detailed analysis based on
 the stability (or better to say, existence) of macroscopic bodies
 permits one to conclude that~$\alpha_L\alt10^{-36}$~\cite{bdov}.

 The effect of screening on the bound~(\ref{E1}) strongly depends on
 the spatial distribution of the source particles in the galaxy. We
 first consider the case where they are normal visible matter so that
 the neutrino path from the SN~1987A, which occurred in the Large
 Magellanic Cloud at a distance of about $50\,\rm kpc$, lies mostly
 outside of the source charge distribution which is concentrated in
 the central region of the Milky Way. In this case the ``Coulomb
 field'' created by this charge along the neutrino path from SN~1987A
 to us is effectively screened by the background neutrinos so that
 practically no bound on their interaction strength can be obtained.

 If the $\nu_e$'s are nearly massless even the cosmic background sea
 remains relativistic today at an effective temperature
 $T_\nu\approx 1.9\,\rm K$. The Debye screening scale in a
 relativistic plasma is by dimensional analysis
 $k_{\rm D}\approx q_\nu T_\nu$ so that a typical screening radius is
 \begin{equation}
 k_{\rm D}^{-1}\approx 3{\times}10^{-23}\,{\rm kpc}/q_\nu.
 \end{equation}
 It is not important that the neutrino phase space distribution is
 precisely thermal. Also, screening occurs even though the background
 neutrinos are a collisionless gas. If $q_\nu\agt10^{-23}$ the
 screening radius is less than $3\,\rm kpc$ and so, the neutrinos
 from SN~1987A would not have felt a force to bend their path. Put
 another way, $q_\nu\agt 10^{-23}$ remains allowed by the SN~1987A
 argument.  If in Eq.~(\ref{E1}) one assumes that $q_{e,p}$ is of the
 same order as $q_\nu$ then there is no meaningful SN~1987A bound at
 all.  What remains excluded is a combination of a very small $q_\nu$
 with a much larger~$q_{e,p}$.

 If the neutrinos are massive so that they are nonrelativistic today
 the screening radius would be even smaller because their total
 number density is fixed at about $100\,\rm cm^{-3}$. If they have
 nonrelativistic velocities they can be polarized even more easily by
 a ``test charge''!

 An interesting bound survives if the source of the new field are
 dark matter particles. Their density distribution in the galaxy
 probably behaves as $r^{-2}$ up to a distance of about $100\,\rm
 kpc$. The equation which governs the screening behavior as a
 function of galactocentric distance can be written as
 \begin{equation}
 dQ_{\nu}/dr=k_{\rm D} (Q_{\rm S}-Q_{\nu})
 \label{E3}
 \end{equation}
 where $Q_{\nu}(r)$ is the charge of the cosmic background neutrinos
 inside the radius $r$ and $Q_{\rm S}(r)$ is the same for the source
 particles. For a localized source charge distribution we have
 $Q_{\rm S}=Q_0$ and the total charge is exponentially screened,
 $Q_{\rm tot}=Q_{\rm S}-Q_{\nu}=Q_0 e^{-k_{\rm D} r}$. If the dark
 matter acts as a source we use $Q_{\rm S}\propto r$ and the total
 charge inside the halo is compensated only with an accuracy
 $(R_0 k_{\rm D})^{-1}$ where $R_0$ is the boundary radius of the
 source charge distribution. The solution of Eq.~(\ref{E3}) then has
 the form
 \begin{equation}
 Q_{\rm tot}(r)=Q_{\rm S}(r)(k_{\rm D} r)^{-1} (1-e^{-k_{\rm D}r}).
 \label{E4}
 \end{equation}
 Here, the factor $Q_{\rm S}(r) (k_{\rm D}r)^{-1}$ is a constant
 because of the assumed behavior $Q_{\rm S}\propto r$.
 Correspondingly, the bound~(\ref{E1}) for dark matter particles now
 takes the form
 \begin{equation}
 q_x < 10^{-17} (m_x/m_p) (r/{\rm kpc}) \approx 10^{-16}\,m_x/m_p.
 \label{E5}
 \end{equation}
 This result does not depend on the neutrino ``charge'' $q_{\nu}$.

 If the background neutrinos are nonrelativistic the Debye length is
 smaller and the bound is weaker. The exact suppression factor
 depends on the neutrino velocity dispersion.

 We have implicitly assumed that the universe is neutral with respect
 to the new charge. If the mass of the corresponding intermediate
 boson is exactly zero this is the only possibility because any
 nonzero cosmic charge density would drastically change cosmology.
 For a nonzero boson mass, even if it is very small, this may not be
 true and a cosmic charge asymmetry is possible. Even in this case
 one would expect screening effects to operate on galactic scales.

 Our argument relies on the assumption of oppositely charged
 neutrinos and antineutrinos, i.e.\ that the new force is mediated by
 vector particles. A force mediated by \hbox{spin-0} or by
 \hbox{spin-2} particles is always attractive. Because there is only
 one conserved \hbox{rank-2} tensor (the energy-momentum tensor) any
 force mediated by massless \hbox{spin-2} particles is identical to
 gravity.  Unless one wishes to consider low-mass bosons as a
 mediator of the novel force, this leaves only the scalar case as a
 realistic option for an unscreened galactic-scale force.

 A scalar force between a static source and relativistic fermions is
 suppressed by a Lorentz factor $m_\nu/E_\nu$. Therefore, in this
 case the bound~(\ref{E1}) is degraded correspondingly; for massless
 neutrinos there would be no bound.  If neutrinos do have a small
 mass, they may partially cluster on galactic scales even though they
 do not contribute the dominant part of the dark matter. In this case
 they may be the dominant source for the force affecting the
 propagation of supernova neutrinos.

 In a recent paper \cite{MST} the dispersion of neutrinos from a
 future supernova near the galactic center was discussed under the
 assumption that the galactic dark matter consists of $\nu_\tau$'s
 with a mass in the $100\,\rm eV$ range. Again, if the force between
 these dark matter neutrinos and those from a SN is mediated by a
 vector force there will be no novel dispersion effect because the
 distribution of galactic dark matter neutrinos is likely CP
 symmetric so that no net charge survives. Even if there were a
 cosmic CP asymmtry of order, say, the baryon asymmetry, the galaxy
 would tend to repel surplus charges to its surface. Within the
 galaxy the ``neutrino plasma'' likely would be neutral.

 {\bf Acknowledgement.} This work was supported, in part, by the
 Theoretical Astrophysics Network (TAN), European Union contract
 CHRX-CT93-0120. A.D.\ acknowledges the hospitality of the
 Max-Planck-Institut f\"ur Physik during a visit when this work has
 been done.


 \end{document}